\documentclass[useAMS,usenatbib]{mn2e}
\usepackage{graphicx}
\usepackage{amssymb, latexsym, amsopn}
\usepackage[T1]{fontenc}

\title[Pressure, oscillations and singularities]{Pressure gradients, shell crossing singularities and acoustic oscillations -- application to inhomogeneous cosmological models}
\author[K. Bolejko and P. Lasky]{K. Bolejko$^{1,2}$\thanks{E-mail:
bolejko@camk.edu.pl,} and P.
Lasky$^{3,4}$\thanks{paul.lasky@sci.monash.edu.au}\\
$^{1}$School of Physics, University of Melbourne, Parkville, Victoria, Australia\\
$^{2}$Nicolaus Copernicus Astronomical Center, Bartycka 18, 00-716 Warsaw, Poland\\
$^{3}$School of Mathematical Sciences, Monash University, Wellington Rd, Melbourne 3800, Australia\\
$^{4}$Theoretical Astrophysics, Eberhard Karls University of T\"ubingen, T\"ubingen 72076, Germany}

\begin{document}


\pagerange{\pageref{firstpage}--\pageref{lastpage}} 

\maketitle

\label{firstpage}

\begin{abstract}
Inhomogeneous cosmological models have recently become 
a very interesting alternative to standard cosmology. This is because these models 
are able to fit cosmological observations without the need for dark energy.
However, due to inhomogeneity and pressure-less matter content, 
these models can suffer from shell crossing singularities.
These singularities occur when two shell of dust collide with each other leading
to infinite values of the density.
In this {\em Letter} we show that if inhomogeneous pressure is included
then these singularities can be 
prevented from occurring over the period of structure formation.
Thus, a simple incorporation of a gradient of pressure allows for more comprehensive studies of inhomogeneous
cosmological models and their application to cosmology.
\end{abstract}


\section{Introduction}
There has been a significant amount of recent interest in inhomogeneous cosmological models as an alternative to the $\Lambda$CDM concordance model [for a review see \citet{C07}].  These inhomogeneous models are able to fit many cosmological observations without the need for dark energy  \citep*{DH98,C00,GSS04,AAG06,AA06,CR06,AA07,EM07,ABNV07,BTT07a,BTT07b,BMN07,MKMR07,KKMA08,BN08,B08,GH08a,E08,YKN08}.  One principle method underlying these schemes is to use inhomogeneous solutions of the Einstein field equations rather than the standard Friedmann-Robertson-Walker solutions used in $\Lambda$CDM models. The implementation of these models suggests that we need to live close to the centre of the Gpc-scale void. There have already been several methods suggested as a test for these
types of models. They are 
based on the time drift of cosmological redshift \citep*{UCE08},
spectral distortions of the CMB power spectrum \citep{CS08},
the kinematic Sunyaev--Zel'dovich effect \citep{GH08b},
future measurements of supernova in the redshift range
of 0.1-0.4 \citep{CFL08} and future Baryon Acoustic Oscillations measurements \citep{BW08}. Based on current observations, the alternative of a Gpc-scale underdensity is indistinguishable from the dark energy scenario.

However, there remain a number of concerns regarding these inhomogeneous cosmologies.  In particular, it is known that pressure-free Lema\^itre--Tolman \citep{L33,T34}
models can evolve to form shell crossing singularities.  This is an additional singularity to the Big Bang that occurs when two shells of matter collide with each other, leading to infinite values of the density.  These singularities were first discussed in the context of astrophysical applications of gravitational collapse, where it has been shown that they can form without being hidden inside an horizon, and are therefore globally naked \citep{YSM73}.  However, they are not considered as real singularities for a number of reasons.  First of all because they are weak in the sense that the spacetime can be extended through the singularity \citep{N86,FK95,N03,PK}.  Secondly, as was shown by \cite{J93}, an object sent through the singularity would not be crushed (i.e. they would not be focused onto a surface or a line).  Finally, in spherically symmetric, asymptotically flat Einstein-Vlasov systems, these types of singularities have been proved not to occur \citep{RRS95}.  However, this has not been proved in cosmological models which are not asymptotically flat.

In gravitational collapse scenario's, shell crossings generally do not require consideration because the initial conditions are generally unrealistic.  However, this is not the case in cosmological models.  \cite{BKH05} showed that a large class of realistic models of voids exhibit shell crossing singularities when the galaxy wall surrounding the void began to form.  In many applications, initial conditions are chosen such that the subsequent evolution does not exhibit shell crossings \citep{HL85}, however this is extremely inconvenient and prevents the study of a wide class of models.

In this {\em Letter} we show that the simple incorporation of pressure gradients leads to an elimination of shell crossings over the time scales involved with structure formation.  Anomalies in the evolution associated with the formation of acoustic oscillations (see Sec. \ref{acousticosc}) prevent us from determining conclusively whether this simple incorporation of inhomogeneous pressure can permanently prevent the shell crossing singularities.  The structure of the {\em Letter} is as follows; In Sec. \ref{ssist} the Lema\^itre model is presented. Sec. \ref{setup} gives an explicit specification of the models being considered.  Sec. \ref{evolution} presents the evolution of both models and shows that sufficiently large pressure gradients can prevent shell crossing singularities. Sec. \ref{acousticosc} then discuss the occurrence of acoustic oscillations due to non-zero pressure gradient.

\section{The spherically symmetric inhomogeneous space--time}\label{ssist}

A spherically symmetric metric in co-moving and synchronous
coordinates can be written as \citep{PK}

\begin{equation}
{\rm d}s^2 = {\rm e}^{A(t,r)} c^2 {\rm d}t^2 - {\rm e}^{B(t,r)}{\rm d}r^2
- R^2(t,r) {\rm d} \Omega^2. \label{ss}
\end{equation}
where ${\rm d} \Omega^2 =  {\rm d}\theta^2 + \sin^2 \theta {\rm d}\phi^2$.
The Einstein field equations for the spherically symmetric perfect
fluid distribution (in coordinate components) are:

\begin{eqnarray}
G^{0}{}_{0} &=& {\rm e}^{-A} \left( \frac{\dot{R}^2}{R^2} + \frac{\dot{B} \dot{R}}{R} \right) - {\rm e}^{-B} \left( 2 \frac{R''}{R} \right. +  \nonumber \\
&+& \left. \frac{R'^2}{R^2} - \frac{B' R'}{R} \right) +
\frac{1}{R^2} = \kappa \epsilon + \Lambda,
\label{G00} \\
G^{1}{}_{0} &=& {\rm e}^{-B} \left( 2 \frac{\dot{R}'}{R} - \frac{\dot{B}
R'}{R} - \frac{A' \dot{R}}{R} \right) =0,
\label{G10} \\
G^{1}{}_{1} &=& {\rm e}^{-A} \left( 2 \frac{\ddot{R}}{R} + \frac{\dot{R}^2}{R^2} - \frac{\dot{A} \dot{R}}{R} \right) + \nonumber \\
&-& {\rm e}^{-B} \left(\frac{R'^2}{R^2} + \frac{A'
R'}{R}\right) + \frac{1}{R^2} = - \kappa p + \Lambda,
\label{G11} \\
G^{2}{}_{2} &=& G^{3}{}_{3} = \frac{1}{4} {\rm e}^{-A} \left( 4
\frac{\ddot{R}}{R}
 - 2 \frac{\dot{A} \dot{R}}{R} + 2 \frac{\dot{B} \dot{R}}{R} +  \right. \nonumber \\
 &+& \left. 2 \ddot{B} +  \dot{B}^2 - \dot{A} \dot{B} \right) - \frac{1}{4} {\rm e}^{-B} \left(4 \frac{R''}{R} + 2 \frac{A' R'}{R} + \right. \nonumber \\
 &-& \left. 2 \frac{B' R'}{R} + 2A'' + A'^2 - A' B' \right) = - \kappa p +
\Lambda, \label{G22}
\end{eqnarray}
where $\epsilon$ is the energy density, $p$ is the pressure, ${\Lambda}$ is the cosmological constant,
$\kappa = 8 \pi G/c^4$,
$\dot{}$ stands for $\partial_t$ and $'$ stands for $\partial_r$.

These equations can be reduced to \citep{L33}
\begin{equation}
 M' = \frac{1}{2} \kappa \rho c^2 R^2 R',~~~\dot{M} = - \frac{1}{2} \kappa p R^2 \dot{R},
\label{mr}
\end{equation}
where $M$ is defined as
\begin{eqnarray}\label{emdef}
&& 2M(r,t) = R(r,t) + R(r,t){\rm e}^{-A(t,r)}{\dot{R}}^2(r,t) - \nonumber \\ 
&& {\rm e}^{-B(t,r)}{R'}^2(r,t)R(r,t) - \frac{1}{3} \Lambda R^3(r,t).
\end{eqnarray}
Although in the literature the above system of equations
is usually credited to \citet{MS64},
it was Lema\^itre who first studied them.
That is why we will refer to the spherically symmetric system with inhomogeneous pressure as the Lema\^itre model.  In the zero pressure case the system will be referred to as the Lema\^itre--Tolman model \citep{L33,T34} (as is widely accepted).

The equations of motion, ${T^{\alpha \beta}}_{;\beta} = 0$, reduce to
\begin{equation}
\dot{B} + 4 \frac{\dot{R}}{R} = - \frac{2 \dot{\rho} c^2}{ \rho c^2 + p},~~~A' = - \frac{2 p'}{ \rho c^2 + p},
\label{Tab}
\end{equation}
Integrating eq. (\ref{G10}) the $B$ function can be expressed as 
\begin{equation}
{\rm e}^{B(r,t)} = \frac{R'^2(r,t)}{1 + 2E(r)} 
 \exp \left( - \int\limits_{t_0}^t {\rm d \tilde{t}} 
\frac{A' \dot{R}}{R'}  \right).
\label{edb}
\end{equation}
where $E(r)$ is an arbitrary function. 
For the zero-pressure Lema\^itre--Tolman model, $A'=0$, implying the time coordinate can be rescaled such that ${\rm e}^{A}=1$ without loss of generality.  Moreover, in this case ${\rm e}^B = R'^2/(1+2E)$.

\section{Model set-up}\label{setup}

\begin{figure*}
\includegraphics[scale=0.65]{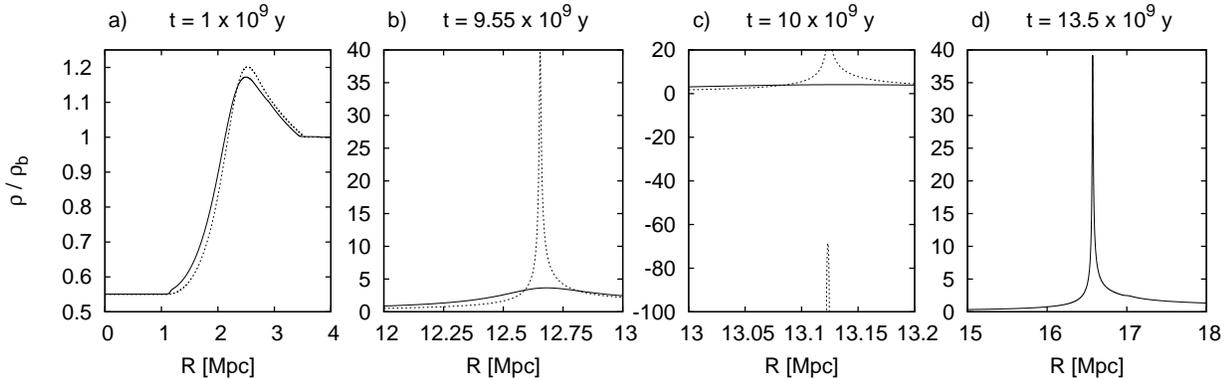}   
\caption{Density distribution for four different instants obtained within the Lema\^itre--Tolman model (dotted line) and Lema\^itre model (solid line). 
The time is indicated at the top of each panel. $R$ corresponds to the value of the areal distance at the given time instant. Both models begin their evolution at the last scattering instant, and they are both defined by the same initial data. At $t = 10^9y$ differences are still small. Between  $t = 9.95 \times 10^9y$
and $t = 10 \times 10^9y$ the shell crossing singularity occurs in the Lema\^itre model.
At this instant, the Lema\^itre--Tolman model breaks down, however we still present
the profile of a function which before the shell crossing was related to the density.
After the shell crossing, $R'<0$ within some region and this leads to unphysical,
negative values of $\rho$, which is depicted in the third panel.}
\label{fig1}
\end{figure*}

To check how the pressure influences the evolution of the spacetime,
we concentrate on two models: the pressure-free Lema\^itre--Tolman model and the Lema\^itre model.
Both models are specified by the same initial data given at the last scattering instant.
The background model is chosen to be a flat Friedmann model with 
$\Omega_{mat} = 0.3$, $\Omega_{\Lambda} =0.7$ and $H_0  = 70$ km s$^{-1}$ Mpc$^{-1}$.
The radial coordinate is defined as areal distance $R$ at the last scattering instant:

\begin{equation}
\tilde{r}:= R(r,t_{ls}).
\end{equation}
However, for clarity in further use,  
the $\tilde{}$ sign is omitted 
and the new radial coordinate is referred to as $r$.
In this {\em Letter} $r$ is always expressed  in kpc.

In this {\em Letter} we are interested in how the gradient of pressure 
influences the evolution of cosmic structure, and if it can prevent the formation of shell crossing singularities.
The most common place where these singularities occur in pressure-free inhomogeneous models
is at the edge of voids.  This is because the 
expansion of space inside underdense regions is larger than inside overdense regions 
and this forces matter to flow  from the voids towards the surrounding galaxy walls, implying concentric shells of matter may collide. 
We therefore consider a model of a void
surrounded by an overdense region.
Within voids, due to the lower amount of matter than in the homogeneous background,
the curvature of the space is negative, thus the explicit
forms  of mass ($M$) and curvature (expressed by the function $E$) are 

\[  M = M_0 + \left\{ \begin{array}{ll}
M_1 r^3 & {\rm ~for~} \ell \leqslant 0.5a, \\
M_2 \exp \left[ - 12 \left( \frac{ \ell - a}{a} \right)^2 \right] & {\rm ~for~} 0.5a \leqslant  \ell \leqslant 1.5a \\ 
M_1 (2a - \ell)^3 & {\rm ~for~} 1.5a \leqslant  \ell \leqslant  2a,  \\
0 &  {\rm ~for~} \ell \geqslant 2a, 
\end{array} \right. \]
\noindent
where $M_0$ is the mass in the corresponding volume of the homogeneous universe [i.e. $M_0 = (4 \pi G /3c^2) \rho_{b,ls} r^3$ and $\rho_{b,ls}$ is the background density at the last scattering instant],
$M_1 = 8 M_2 a^{-3} {\rm e}^{-3/2}$, $M_2 = -0.3$ kpc, $a = 12$ kpc.

\[ E = - \frac{1}{2} \times  \left\{ \begin{array}{ll}
E_1 r^2 & {\rm ~for~} r \leqslant 0.5 b, \\
E_2 \exp \left[ - 4 \left( \frac{ r - b}{b} \right)^2 \right] & {\rm ~for~} 0.5 b \leqslant  r \leqslant 1.5 b \\ 
E_1 (2b - r)^2 & {\rm ~for~} 1.5 b \leqslant  r \leqslant  2 b,  \\
0 &  {\rm ~for~} r \geqslant 2b, 
\end{array} \right. \]
\noindent
where $E_1 = 4 E_2 b^{-2} {\rm e}^{-1}$, 
$E_2 = -1.1 \times 10^{-5}$, $b = 10.9$ kpc.
It should be noted that other models of voids are also possible --
even ones which do not evolve from initial 
rarefactions but from condensation, cf. \citet{MH01}.
However, this particular  void model was chosen because it develops, as we will 
show, a shell crossing singularity.
As can bee seen for $r>24$ kpc\footnote{For the current instant this corresponds to areal distance of $R \approx 26$ Mpc.}
 the mass distribution as well as the curvature is the same 
as in the homogeneous FLRW models.
These functions were used as an initial condition specified
at the last scattering instant. The initial density distribution for these models is very close to the form given in the first panel of Fig \ref{fig1}.  One can see here that the void region extends from $R\approx 1.5$ Mpc\footnote{This value of the areal radius at $t= 10^9 y$ corresponds to a present value of around 16.5 Mpc.}, and is surrounded by the galaxy wall which has a density up to twice the value of the void.

We start the evolution of both the pressure-free Lema\^itre--Tolman
and the Lema\^itre models from the same profile of mass and curvature distributions.
The only discrepancy between these models is with the equation of state,
which was chosen to be of a polytropic form
\begin{equation}
p = K  \rho^{1 + 1/n}.
\label{peos}
\end{equation}
The polytropic index is chosen to be $n =3/2$ which is the case of a mono-atomic gas.
This equation of state is a good approximation
to describe degenerate star cores,  giant gaseous planets, or even for rocky planets.
Thus, although realistic conditions within high-density regions inside walls
 might lead to a more complicated dependence of pressure,  this simple polytropic equation of state can be treated as a good first approximation to the problem considered in this {\em Letter}. The constant K for the Lema\^itre--Tolman model, which is pressure-free, is $K=0$
and for Lema\^itre model is chosen to be  $K = 1.98 \times 10^{14}$ m$^2$/s$^2$.
(Sec. \ref{evolution}) and $K = 1.08 \times 10^{14}$ m$^2$/s$^2$
(Sec. \ref{acousticosc}).
These are very high values. For comparison 
the ratio of standard pressure of air ($p_{air} = 101.325$ kPa)
to its density ($\rho_{air} = 1.292$ kg/m$^3$)
at T=0$^{\circ}$C is approximately equal to $7.84 \times 10^4$ m$^2$/s$^2$.
Such values were chosen in order to better depict the influence 
of pressure gradients on the evolution of matter. However, even if such 
very stiff equations of state are employed, their impact on the evolution is visible only 
when density gradients become large. Thus, the incorporation of this gradient of pressure mostly affects only regions where the shell crossing singularities would occur.

The algorithm which is used to calculate the evolution in the Lema\^itre--Tolman model is the same as the one used and described in \citet{BKH05} and 
the algorithm which is used to calculated the evolution 
in the Lema\^itre model is the same as the one used 
and described in \citet{B06}.

\section{Evolution}\label{evolution}

\begin{figure}
\includegraphics[scale=0.65]{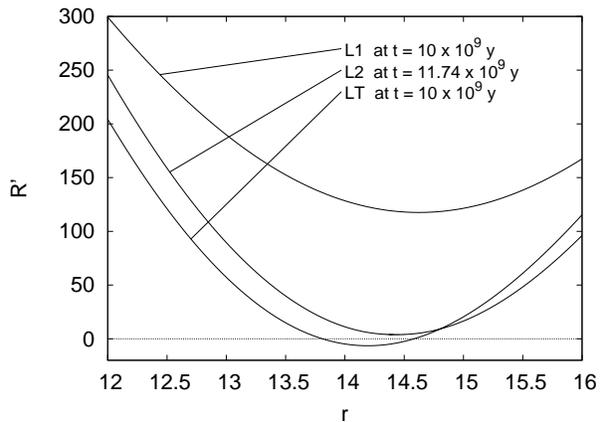}   
\caption{The dependence of $R'(r)$. The curve denoted as L1 corresponds to the
value of $R'$ obtained at $t= 10 \times 10^9$ y within the Lema\^itre model presented in Sec. \ref{evolution} (see Fig. \ref{fig1}c).
Curve denoted as L2 corresponds to the
value of $R'$ obtained at $t= 11.74 \times 10^9$ y within the Lema\^itre model considered in Sec. \ref{acousticosc} (see Fig. \ref{fig3}).
Curve denoted as LT corresponds to the
value of $R'$ obtained at $t= 10 \times 10^9$ y within the Lema\^itre--Tolman model (see Fig. \ref{fig1}c).}
\label{fig2}
\end{figure}

The evolution of matter in both the Lema\^itre and Lema\^itre--Tolman models is presented in Fig. \ref{fig1}.
Both models start from the same initial conditions at the last scattering instant.
As can be seen at $t = 10^9$y (Fig. \ref{fig1}a) the difference between density distributions in these models 
is still small --  this is because the pressure gradients are still small.\footnote{It should be 
noted that it is the gradient of pressure that matters. In the case of pressure being only a function of time, the function ${\rm e}^B$ has the same form as in the Lema\^itre--Tolman model,
and the evolution of the system with pressure is very similar to the evolution 
without pressure -- see \citet{BKH05} for details.}
However, with subsequent evolution the density contrast increases, and hence the pressure gradient also increases.
At $t = 9.95 \times 10^9$y (Fig. \ref{fig1}b) the difference 
between the evolution of matter within the models with and without pressure 
is significantly large.
The ratio of $\rho/\rho_b$ at the maximum in the Lema\^itre--Tolman model is almost 40, whereas in the Lema\^itre model it is around 3.7.
At this instant in the Lema\^itre--Tolman model, 
for $r$ close to the maximum, $R'$ is almost zero.
Soon, the shell crossing occurs ($R'=0$, $M'\ne 0$).
At $t = 10 \times 10^9$y  (Fig. \ref{fig1}c) the  density in the Lema\^itre model is everywhere finite -- 
at the maximum $\rho/\rho_b \approx 4.1$. However,
the Lema\^itre--Tolman model is not applicable, since it breaks down at the shell crossing.
This is because after the shell crossing, $R'<0$ (see curve LT in Fig. \ref{fig2}), implying from equation (\ref{mr}) that the density becomes negative. 
At the current instant, the density in the Lema\^itre model is still finite
and free from shell crossings.
This shows that the sufficiently large pressure gradient can prevent
shell crossing singularities over the period of structure formation and 
their evolution.
However as it will be shown in the next section it is not certain if
simple inhomogeneous pressure can permanently prevent the occurrence of these singularities.

\section{Acoustic oscillations}\label{acousticosc}

\begin{figure}
\includegraphics[scale=0.65]{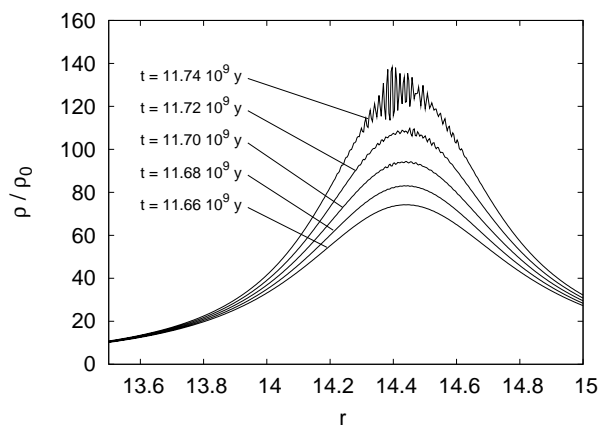}   
\caption{Density distribution for five different instants for a model with equation
of state $p =  1.08 \times 10^{14}$ $\rho^{5/3}$ [Pa] (see  Sec. \ref{acousticosc} 
for details).}
\label{fig3}
\end{figure}

In this section, the constant $K$ is almost half of the value used in the previous section.
For smaller values of pressure, the density increases faster than
previously which leads to a larger density contrast, which subsequently leads to larger pressure gradients.
This causes oscillations which are  depicted in Fig. \ref{fig3}.
The nature of these oscillations is visible in the following equation
[this is a rearranged eq. (\ref{emdef})]

\begin{eqnarray}
&& {\rm e}^{-A} \dot{R}^2 = \frac{2M}{R} + \frac{1}{3} \Lambda R^2 +  \nonumber \\
&& (1+2E) \exp \left(- 2 \int {\rm d} t \frac{p'}{\rho c^2 + p} \frac{\dot{R}}{R'} \right) - 1.
\end{eqnarray}
As seen from the above expression, when in a region where the gradient of pressure is too large, the expansion rate, ${\rm e}^{-A} \dot{R}^2$, slows down.
When one shell slows, then a shell with larger $r$ expands relatively faster and the pressure gradient
drops, which leads to change of sign of $p'$ (initially
$p'$ was an increasing function from $0$ to $r_{max}$ -- where $r_{max}$
correspond to the maximum).
When $p'$ changes sign, the expansion rate increases again. This is the origin of the oscillations.
However, at one stage these oscillations become so large that $\dot{R}$ can change
sign and some shells start to collapse. 
Within the model considered in this section, the first shells
of matter start to collapse at $t=  11.76 \times 10^{9}$y
(note that the oscillations presented in Fig. \ref{fig3}
are shown up to $t=  11.74 \times 10^{9}$y)
At this stage the oscillations are so rapid and of large amplitude
that numerical modelling of this phenomena becomes very difficult.
This prevents us from further studying the evolution and to check
if the gradient of pressure is sufficient to prevent the shell crossing.
The value of $R'$ at  $t=  11.74 \times 10^{9}$y is presented in Fig. \ref{fig2}
(curve L2). 

We also proceeded with smaller values of the spatial and time steps. However, after $\dot{R}<0$ the
analysis becomes too difficult to handle properly within the comoving and synchronous gauge (\ref{ss}), due to the steep gradients in the density.
This suggests that a more comprehensive and detailed study of shell
crossings singularities and acoustic oscillations should be performed within
other coordinate systems. For example 
the Gautreau coordinates for perfect fluid spacetimes  discussed in detail and studied by \citet{LL07a,LL07b}
can provide a more suitable insight. 
The Gautreau coordinates have already proved their usefullness in studying
shell crossing singularities for dust spacetimes  [see for example \citet*{N86,FK95,N03,PK,LLB07}].

\section{Conclusions}\label{concl}

In this {\em Letter} we studied the shell crossing singularities which occur in pressure-free 
spherically symmetric cosmological models.
These models have recently become very popular because they
can fit cosmological observations without the need for dark energy.
However, because of shell crossing singularities, which occur when two shells
of matter collide, the full parameter space cannot be considered.
We have investigated the incorporation of pressure into these models. 
Adding a gradient of pressure implies the equations can no longer be solved analytically, although conceptually they are no more difficult -- see eq. (\ref{mr}) -- and they are still simple to compute numerically. 
We showed that the incorporation of a pressure gradient 
can prevent shell crossing singularities occurring over the period of structure formation.
This will enable the unlimited investigation of these inhomogeneous cosmological models without the occurrence of these anomalous singularities. 

We also showed in this {\em Letter} that,
in some cases, when the density contrast is sufficiently high,
the existence of pressure gradients leads to acoustic
oscillations. At this stage the numerical investigation
becomes troublesome.
However, by choosing a harder form of the equation of state we can postpone the occurrence of shell
crossing singularities so they do not occur  over the time of structure evolution. 
It should be noted that such harder equations of state
require that the constant $K$ in (\ref{peos}) be unnaturally large -- by several orders of magnitude larger that the realistic value. This means that to fully describe the process of acoustic oscillations
we need to perform a more comprehensive analysis.
In particular, an application of more realistic fluids which allows for an anisotropic pressure and heat flow is essential.

\section*{Acknowledgments}

KB would like to thank Peter and Patricia Gruber 
and the International Astronomical Union for the PPGF Fellowship.  PL was partially supported by the DFG via SFB/TR7.

\label{lastpage}

\end{document}